%% file: main.tex
\documentclass{article}

\PassOptionsToPackage{numbers,square}{natbib}
\usepackage[preprint]{neurips_2026}
\usepackage[utf8]{inputenc}
\usepackage[T1]{fontenc}
\usepackage{amsmath,amssymb,mathtools}
\usepackage{algorithm}
\usepackage[noend]{algpseudocode}
\usepackage{booktabs}
\usepackage{graphicx}
\usepackage{xspace}
\usepackage[table]{xcolor}
\usepackage{colortbl}
\usepackage{tikz}
\usepackage{tcolorbox}
\tcbuselibrary{skins}
\usepackage{titlesec}
\usepackage[labelfont=bf,font=small,labelsep=period]{caption}
\usepackage{microtype}
\usepackage{hyperref}
\usepackage{url}

\definecolor{DeepIndigo}{RGB}{10,20,100}
\definecolor{ElectricCyan}{RGB}{0,220,255}
\definecolor{caspianbg}{RGB}{248,250,255}
\hypersetup{
  colorlinks=true,
  linkcolor=DeepIndigo,
  citecolor=DeepIndigo,
  urlcolor=ElectricCyan,
  pdftitle={Origin Attribution for Unsafe Behavior in Self-Evolving Multi-Agent Coding Systems},
  pdfauthor={Renxiang Wang and Jiaming Cui}
}
\titleformat{\section}
  {\normalfont\Large\bfseries\color{DeepIndigo}}{\thesection}{1em}{}
\titleformat{\subsection}
  {\normalfont\large\bfseries\color{DeepIndigo!80}}{\thesubsection}{1em}{}

\definecolor{sigcontent}{HTML}{0072B2}
\definecolor{sigtiming}{HTML}{009E73}
\definecolor{rankbest}{HTML}{C8DDF0}
\definecolor{ranksecond}{HTML}{EBF2F8}

\input{numbers}

\AddToHook{shipout/before}{%
  \pdfpageattr{/Group<</S/Transparency /I true /CS/DeviceRGB>>}%
}

\renewenvironment{abstract}{%
  \begin{center}
  \begin{tcolorbox}[
    enhanced,
    colback=caspianbg,
    colframe=gray!5,
    arc=0mm,
    outer arc=0mm,
    width=\linewidth,
    left=4mm,
    right=4mm,
    top=3mm,
    bottom=3mm,
    boxrule=0pt,
    borderline west={2.5pt}{0pt}{DeepIndigo!90!ElectricCyan}, 
    before skip=10pt,
    after skip=20pt
]
  \noindent\ignorespaces
}{%
  \par\end{tcolorbox}
  \end{center}
  \vspace{0.1cm}
}

\begin{document}

\begin{center}
    \vspace*{-1.2cm}
    \begin{minipage}[c]{0.15\linewidth}
        \includegraphics[height=0.85cm]{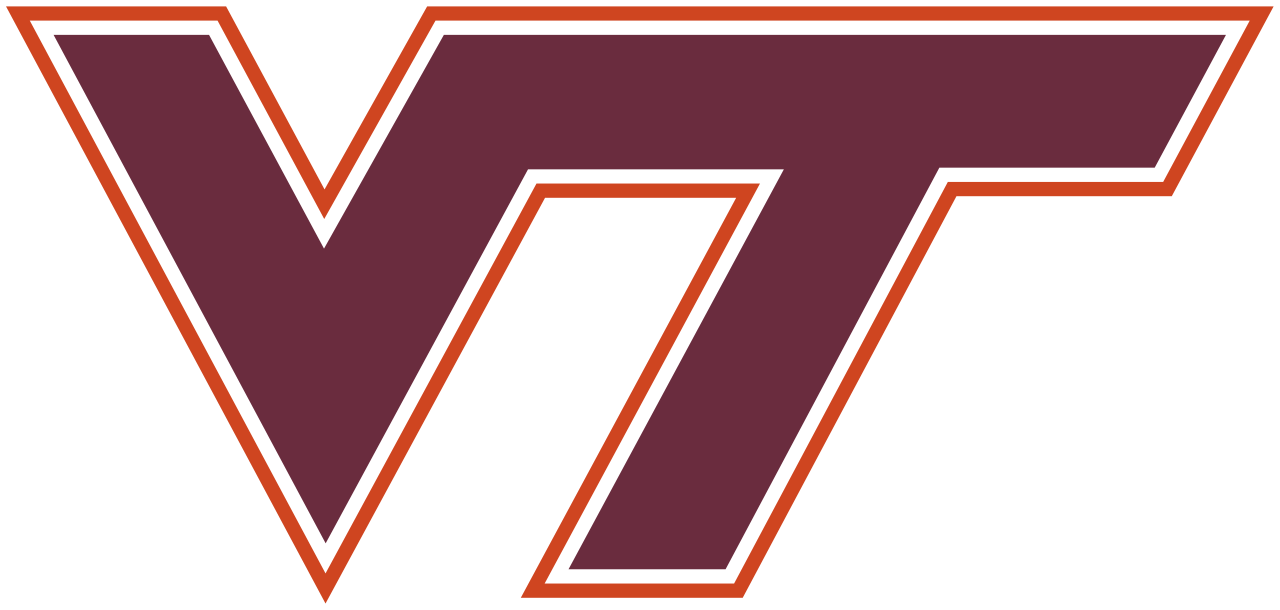}
    \end{minipage}%
    \begin{minipage}[c]{0.15\linewidth}
        \includegraphics[height=0.85cm]{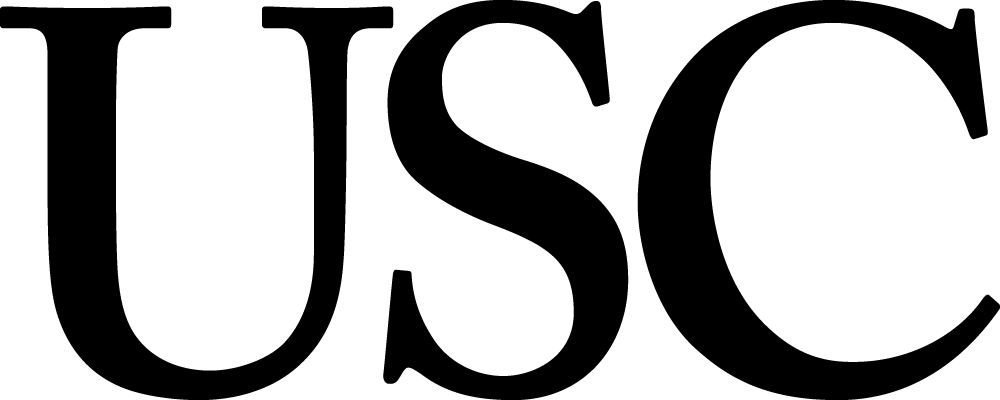}
    \end{minipage}%
    \hspace{-1cm}
    \begin{minipage}[c]{0.5\linewidth}
        \centering
        \footnotesize\textcolor{gray!70}{Preprint \\ \today }
    \end{minipage}%
    \begin{minipage}[c]{0.25\linewidth}
        \hfill 
    \end{minipage}
    
    \vspace{0.25cm}
    {\color{DeepIndigo!20}\rule{\linewidth}{0.6pt}}
    \vspace{0.4cm}
\end{center}
\begin{center}
{\LARGE \bfseries
Origin Attribution for Unsafe Behavior  in Self-Evolving Multi-Agent Coding Systems\\
\par}

\vspace{0.25cm}

{\large
Hanyi Yu$^{1}$ \quad
Jiaming Cui$^{2}$ \\}
\vspace{0.05cm}
{$^{1}$\texttt{University of Southern California, Los Angeles, CA}\\}
{$^{2}$\texttt{Virginia Tech, Blacksburg, VA}\\}
{\small
\texttt{hanyiyu@usc.edu, jiamingcui@vt.edu}\\}

\vspace{0.15cm}

\vspace{0.5cm}
\end{center}

\input{sections/01_intro}
\input{sections/07_related}
\input{sections/02_setting}
\input{sections/03_env}
\input{sections/04_method}
\input{sections/05_experiments}
\input{sections/06_discussion}
\input{sections/08_conclusion}
\input{sections/09_statements}

{\small
\bibliographystyle{plainnat}
\bibliography{refs}
}

\clearpage
\appendix
\input{sections/A_appendix}

\end{document}

%% file: numbers.tex
\newcommand{\NRows}{158\xspace}              
\newcommand{\NPos}{135\xspace}               
\newcommand{\NNeg}{23\xspace}                
\newcommand{\NScenes}{85\xspace}             
\newcommand{\MaxAgents}{7\xspace}            
\newcommand{\Kbudget}{3\xspace}              

\newcommand{\AccText}{20.9\%\xspace}         

\newcommand{\AccTextDopen}{5.1\%\xspace}     

\newcommand{\NegOurs}{100\%\xspace}

\newcommand{\StackThreeOurs}{90\%\xspace}    

\newcommand{\AblTransport}{26.6\%\xspace}    
\newcommand{\AblThree}{75.3\%\xspace}        
\newcommand{\AblNoEarliest}{27.8\%\xspace}   
\newcommand{\AblNoPrintPos}{84.4\%\xspace}     
\newcommand{\AblThreePos}{71.1\%\xspace}
\newcommand{\AblNoPrintNeg}{0\%\xspace}      
\newcommand{\AblShipped}{96.8\%\xspace}      

\newcommand{\ExecOurs}{59.3\xspace}
\newcommand{\LatencyOurs}{0.57\xspace}       

\newcommand{\LLMCore}{15.2\%\xspace}         

\newcommand{\AccBudgetFive}{97.5\%\xspace} 

\newcommand{\AccExhaustiveMatch}{77.2\%\xspace} 

%% file: sections/01_intro.tex
\begin{abstract}
As language-model-based coding agents evolve their programs and reuse code from one another, safety failures can acquire a history that extends beyond the agent in which they are detected. An unsafe change may originate in one agent, propagate through intermediate recipients, and survive subsequent revisions before an audit reveals it. Understanding this history is important for identifying the introducing change and distinguishing inherited behavior from independently occurring failures. We study origin attribution in self-evolving multi-agent coding systems: recovering the agent and program version that introduced an observed unsafe behavior into the target's code lineage, or determining that it arose locally. This task is challenging because rewriting obscures textual similarity, independent occurrences confound temporal ordering, and intermediate recipients can resemble the source. We propose a method that combines behavioral fingerprint matching with historical localization and recursive tracing. It compares candidate implementations beyond the unsafe outcome, retains evidence from multiple historical versions, and follows matching candidates upstream. On \NRows{} queries across \NScenes{} controlled scenarios, the method achieves \AblShipped{} exact attribution accuracy, compared with \AccExhaustiveMatch{} for exhaustive fingerprint matching without recursion. Component ablations and evaluation on 34 additional queries using the same method and hyperparameters demonstrate the complementary roles of multiple representatives and recursive tracing. Further checks extend the evaluation to approval-boundary and resource-owner errors. Together, these results show how behavioral and historical evidence can recover the origins of unsafe behavior through code revision and cross-agent propagation. Code is available \href{https://anonymous.4open.science/r/origin-in-self-evovling-multi-agent-8818/README.md}{\textcolor{blue}{here}}.
\end{abstract}

\section{Introduction}
\label{sec:intro}

Language-model-based coding agents increasingly revise their own programs and reuse code from other agents. These capabilities can also preserve and spread unsafe behavior. Studies of self-evolving agents show that successful task performance can coexist with safety degradation \citep{misevolution,selfevolvesafety}: a customer-service agent, for example, approved an improper refund, received a high rating, and subsequently treated refunds as a way to earn high ratings \citep{misevolution}. Unsafe experience may persist as reusable procedures \citep{skillmisevolution}, while cross-agent sharing can propagate compromised components through a population \citep{selfevolvingsafety}.

Once an audit detects such behavior, identifying the affected agent is only the starting point. The introducing change may lie in another agent's history and may have passed through several recipients before reaching the audited version. Prior work calls for attributing safety degradation to specific evolutionary events \citep{selfevolvingsafety}, and shared-memory research identifies lost artifact provenance as an obstacle to tracing reuse \citep{governedmemory}. We therefore ask: \emph{which agent, at which version, introduced the unsafe behavior into the target's code lineage?}

Answering this question requires distinguishing inheritance from resemblance. Rewriting can obscure textual similarity, while independently introduced failures can share an unsafe outcome. Code-completion studies also show that generated code can reproduce historical bugs \citep{bugreplicators}. Even a matching donor may be an intermediate recipient rather than the origin. These ambiguities connect agent failure attribution \citep{whoandwhen,agenttracer} with software-history analysis \citep{szz}, but arise across persistent code-adoption histories rather than a single execution trace.

Our method follows this historical structure. It anchors the audit at an early unsafe target version, compares candidate code on behavior beyond the failure itself, and traces matching candidates toward earlier versions in other agents. Multiple representatives retain evidence from both unsafe onset and later revisions. Behavioral compatibility identifies plausible candidates, while historical localization and recursive tracing seek the introducing event behind later revisions and intermediate recipients. The method uses archived code and timestamps, with adoption records reserved for evaluation.

We contribute an origin-attribution task and controlled benchmark, together with a method for tracing unsafe behavior through revision and adoption. The evaluation systematically varies rewriting, independent occurrences, and propagation, and shows how behavioral matching and historical tracing jointly improve source recovery.

%% file: sections/07_related.tex
\section{Related Work}
\label{sec:related}

\begin{figure}[t]
\centering
\includegraphics[width=\linewidth]{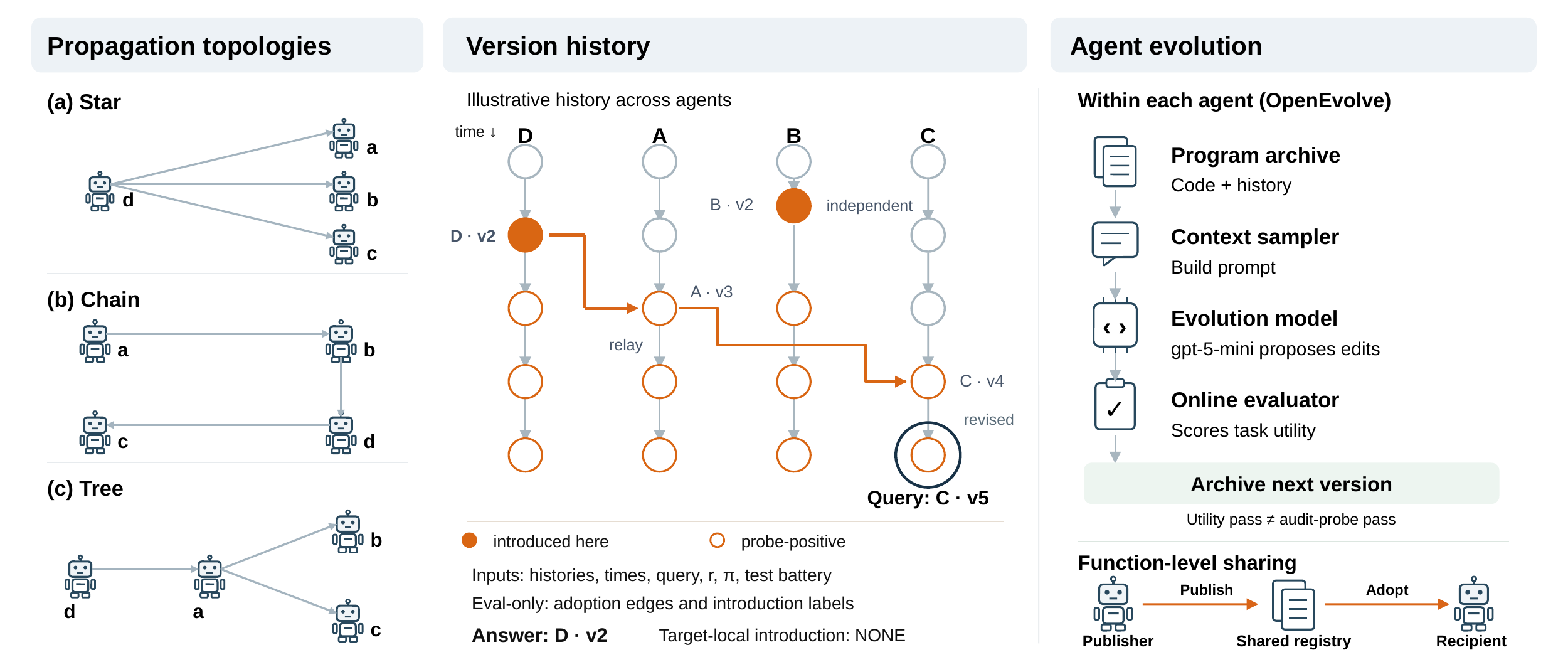}
\caption{Environment and attribution task (illustrative history). $D\!\cdot\!v_2$ propagates through $A\!\cdot\!v_3$ to $C\!\cdot\!v_4$; $C\!\cdot\!v_5$ is audited after further revision. $B\!\cdot\!v_2$ is an earlier independent occurrence. Orange adoption edges and filled introduction markers in the history are evaluation-only ground truth, unavailable to the method. Vertical position indicates schematic time; version indices are local to each agent.}
\label{fig:benchmark}
\end{figure}

\paragraph{Agent failure attribution.}
Who\&When identifies responsible agents and execution steps \citep{whoandwhen}. AgenTracer uses counterfactual replay and fault injection to construct training trajectories \citep{agenttracer}; VerifyMAS and TraceElephant further study trace-level attribution \citep{verifymas,traceelephant}. MAST categorizes recurring failures, while CASPIAN studies cross-agent cascades within a session \citep{mast,caspian}. Our attribution unit is an introducing agent--version pair across persistent code-adoption histories. The executing agent, immediate donor, and origin can differ.

\paragraph{Intervention and controlled evaluation.}
DoVer evaluates attribution hypotheses by measuring recovery after message and plan interventions \citep{dover}; MP-Bench examines multiple defensible attribution perspectives \citep{rethinkingattribution}. We instead evaluate recovery of one recorded introducing event per queried code lineage. AgentChaos, MAS-FIRE, and AEGIS inject or generate faults for agent evaluation \citep{agentchaos,masfire,aegis}, while AgentTrace uses causal graph tracing \citep{agenttrace}. Our controlled changes persist through code revision and adoption; recorded transfers establish labels but remain hidden from attribution methods.

\paragraph{Persistent safety risks.}
Voyager illustrates executable-skill reuse across tasks \citep{voyager}. Misevolution, experience-driven safety evaluations, and Skill Misevolution examine degradation during evolution and its persistence in reusable procedures \citep{misevolution,selfevolvesafety,skillmisevolution}. Cross-generational safety analyses describe risks spreading across agent populations \citep{selfevolvingsafety}, while governed shared memory emphasizes explicit provenance \citep{governedmemory}. Our task asks how far archived code and timestamps support retrospective recovery when such derivation links are unavailable.

\paragraph{Software history and fault localization.}
SZZ links fixes to defect-introducing changes \citep{szz}; clone detection and file-copying studies investigate similarity and reuse \citep{sourcerercc,nicad,filelevelcopy}. Our audit starts from an unsafe version without requiring a fix. Delta debugging and mutation-based fault localization use interventions and execution outcomes \citep{deltadebugging,muse,metallaxis}. We test historical code in the target context, compare behavioral fingerprints, and combine temporal selection with upstream tracing. Rewriting can obscure textual reuse, while independent failures mislead onset-based selection.

%% file: sections/02_setting.tex
\section{Problem Formulation}
\label{sec:setting}

We consider coding agents that revise programs and adopt code from one another, motivated by executable-skill reuse \citep{voyager} and cross-agent inheritance \citep{selfevolvingsafety}. Each agent $a$ maintains an archive $H_a=(a_0,\ldots,a_{n_a})$, which may include branching local revisions. For function-level adoption, write $p|_r$ for function $r$ in program $p$. An edge $(a,a_i)\rightarrow(b,b_j)$ records that $b_j$ incorporated region $r$ from $a_i$; subsequent revisions may modify that code.

\paragraph{Audit query and available evidence.}
An audit query specifies a target $(t,t_q)$, a deterministic probe $\pi$ with $\pi(t_q)=1$, and a function $r$. The auditor receives the agents' collected code histories and timestamps, together with a fixed task-specific input battery $B$ for behavioral comparison. We study retrospective attribution when transfer-level provenance is unavailable, in contrast to systems that explicitly record derivation links \citep{governedmemory}. Adoption records and injection labels are reserved for evaluation. The audit probe is independent of the online evaluator used during evolution. In an audit workflow, the observed failure supplies a reproducible probe, and the audit identifies the affected version and relevant function. Attribution then searches the archived histories using these inputs.

\paragraph{Historical origin.}
The desired output is the introducing agent--version pair $(a^\ast,v^\ast)$ on the target's derivation lineage, tracing through intermediate recipients. If the behavior arose within the target without an external source, the answer is \textsc{None}. These are negative examples for external-source attribution even though the target is unsafe. Evaluation requires an exact pair match or the correct no-source answer.

Each queried lineage has one introducing event; unrelated branches and other agents may independently exhibit the same behavior. The recorded lineage therefore determines a unique evaluation label, but the auditor's observations need not uniquely identify it. The next section describes how we construct histories with known origins while withholding their derivation records from the attribution methods.

%% file: sections/03_env.tex
\section{The Evolving Multi-Agent Coding Environment}
\label{sec:env}

Our environment uses OpenEvolve as the base evolutionary coding framework for each agent \citep{openevolve}. A shared registry supports code publication and adoption under star, chain, and tree organizations drawn from MARBLE \citep{marble}. Figure~\ref{fig:benchmark} shows the environment and an illustrative origin-attribution query.

\paragraph{OpenEvolve-based agents.}
Each agent starts from a common initial program and runs its own OpenEvolve instance. The evolution model, \texttt{gpt-5-mini}, proposes revisions; an online evaluator scores task utility, and an archive retains versions. Local evolution continues between publication and adoption events. The task is workflow routing. The unsafe change grants a designated request kind special routing with confirmation disabled, bypassing the default handling of unknown requests. Appendix~\ref{app:construction} specifies the models, schedules, and evaluation inputs.

\paragraph{Communication and propagation.}
Agents publish versions and adopt functions through the registry. A star connects a source to several recipients; a chain passes code through successive agents; a tree branches to multiple descendants. Cyclic scenarios exchange benign code while the unsafe behavior originates within the target. Topology names describe the unsafe propagation backbone; unrelated adoption can add other edges. Each transfer records the source version, recipient version, and region for evaluation labels.

\paragraph{Scenario construction.}
Following controlled fault-injection approaches for agent evaluation \citep{agentchaos,aegis}, we introduce an unsafe change at a designated program version and track its subsequent adoption. The change passes the online evaluator, while a separate audit probe detects the routing bypass. This separates when a change is introduced from when an audit reveals it. Each queried propagation lineage has a known introducing event. Scenarios without an external source instead introduce the behavior within the target. The fixed audit probe is excluded from online selection and generation prompts; origin labels and adoption logs are reserved for evaluation.

To evaluate attribution under code reuse and revision, we vary semantic rewriting (\textbf{A}), independent co-occurrence (\textbf{B}), unrelated adoption (\textbf{C}), and multi-hop propagation (\textbf{E}). In condition B, we place an unrelated occurrence before the true source to test whether attribution distinguishes temporal precedence from derivation. The generator combines these conditions to evaluate their joint effects. Evidence drift (\textbf{D}) is evaluated on the same histories by querying the target's last flagged version instead of its first. Table~\ref{tab:modes} summarizes the conditions.

Across these conditions, the generators vary evolution schedules and agent start order while withholding change descriptions from attribution inputs. Section~\ref{sec:setup} describes the resulting evaluation set.

%% file: sections/04_method.tex
\section{Method}
\label{sec:method}

\begin{figure}[t]
\centering
\includegraphics[width=\linewidth]{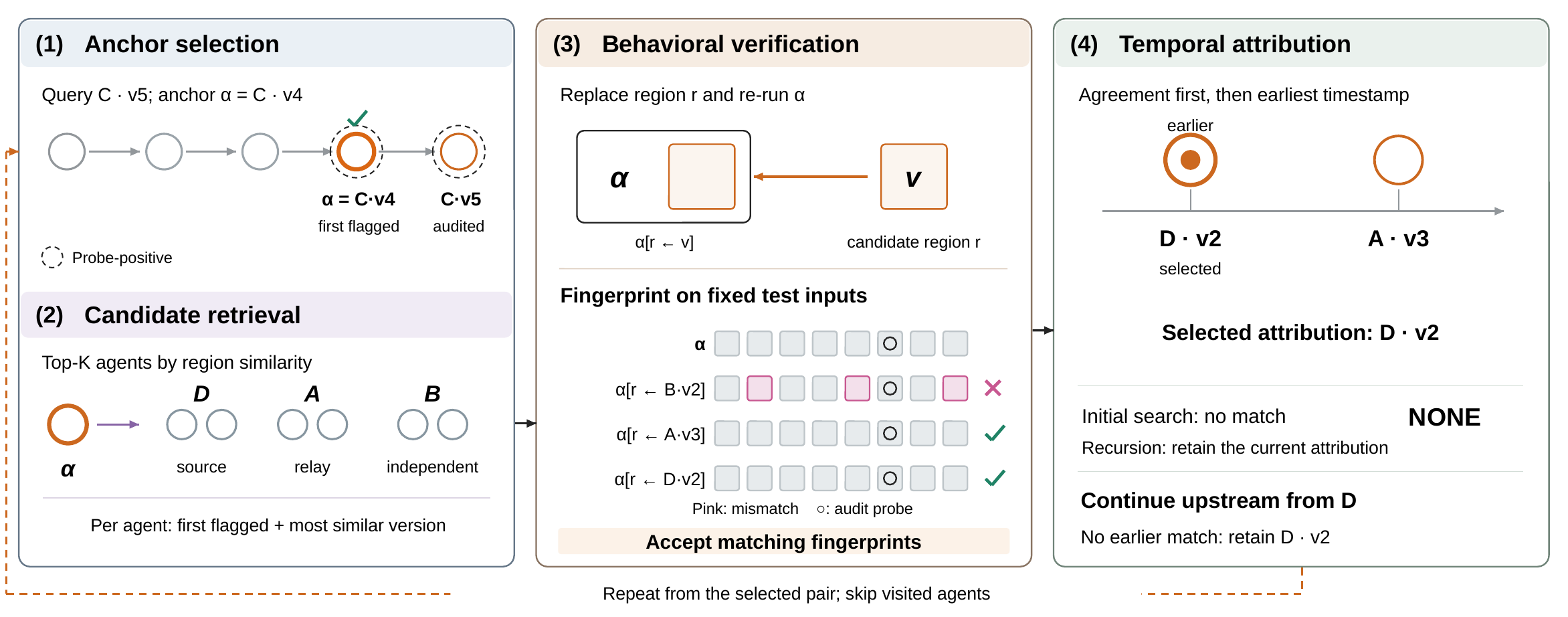}
\caption{Attribution on the example in Figure~\ref{fig:benchmark}. The matrix illustrates one replacement per candidate agent: $B\!\cdot\!v_2$ disagrees with the anchor beyond the unsafe outcome, while $A\!\cdot\!v_3$ and $D\!\cdot\!v_2$ match. Agreement ties favor the earlier timestamp, selecting $D\!\cdot\!v_2$ here. Recursive search retains the current attribution when no earlier match is found; an empty accepted set at the initial tracing level returns \textsc{None}.}
\label{fig:method}
\end{figure}

The method anchors an unsafe target version, matches historical candidates, and traces them upstream. Consider $A\rightarrow B\rightarrow C$: $B$ first adopts the unsafe code from $A$, then revises routine behavior before passing it to $C$. The later version of $B$ can match $C$ even when its onset version and $A$ do not. Multiple representatives expose this later match; recursion returns to $B$'s onset, where the earlier match to $A$ can remain intact. With only the onset representative, tracing may miss $B$; without recursion, it stops at $B$. Figure~\ref{fig:method} illustrates the matching and selection steps.

\paragraph{Historical localization and candidate retrieval.}
Ordering each history $H_a$ by creation time, we locate the first flagged version up to the audited target $t_q$:
\begin{equation}
\iota_q(t)=\min\{i\leq q:\pi(t_i)=1\},
\qquad \alpha=t_{\iota_q(t)}.
\label{eq:anchor}
\end{equation}
Using this historical anchor makes the comparison independent of subsequent target revisions.

For each other agent $a$, we retrieve the version whose code region $r$ is most similar to the anchor and use its similarity as the agent's retrieval score:
\begin{equation}
\begin{aligned}
v_{\mathrm{sim}}(a;\alpha)
  &= \operatorname*{arg\,max}_{v\in H_a}\operatorname{sim}(v|_r,\alpha|_r),\\
s(a;\alpha)
  &= \operatorname{sim}\bigl(v_{\mathrm{sim}}(a;\alpha)|_r,\alpha|_r\bigr).
\end{aligned}
\label{eq:retrieval-score}
\end{equation}
Let $V$ be the set of agents already visited during tracing, initially empty. We retain the top-$K$ agents, excluding the target and previously visited agents:
\begin{equation}
C_K(\alpha,V)=\operatorname*{TopK}_{a\notin V\cup\{t\}}s(a;\alpha).
\label{eq:recall}
\end{equation}
Each shortlisted agent contributes up to two representative versions. Writing $v_{\mathrm{on}}(a)$ for the first version in $H_a$ flagged by $\pi$, when one exists, we define
\begin{equation}
R(a;\alpha)=\begin{cases}
\{v_{\mathrm{on}}(a),v_{\mathrm{sim}}(a;\alpha)\}, & \text{if a flagged version exists},\\
\{v_{\mathrm{sim}}(a;\alpha)\}, & \text{otherwise}.
\end{cases}
\label{eq:representatives}
\end{equation}
The onset representative captures the initial unsafe implementation; the most similar representative can retain evidence of a later revision adopted by the target. Coincident representatives are tested once.

\paragraph{Candidate replacement and behavioral matching.}
For each representative $v$, we replace region $r$ in the anchor with the corresponding candidate code, yielding the hybrid program $\alpha[r\leftarrow v]$. Following intervention-and-reexecution techniques \citep{deltadebugging,muse,metallaxis}, we test whether this substitution preserves the anchor's observed behavior.

Let $B$ be a fixed battery of inputs and $g(p,x)$ the observable response signature of program $p$ on input $x$. We define its behavioral fingerprint as
\begin{equation}
\phi_B(p)=\left(\bigl(g(p,x)\bigr)_{x\in B},\,h_\pi(p)\right),
\label{eq:fingerprint}
\end{equation}
where $h_\pi(p)$ records the audit probe's outcome. The signatures record tool selection, execution mode, and confirmation requirements, comparing behavior beyond the unsafe outcome.

The accepted candidate set is
\begin{equation}
S(\alpha,V)=\left\{(a,v):
\begin{array}{l}
a\in C_K(\alpha,V),\quad v\in R(a;\alpha),\\
\phi_B\bigl(\alpha[r\leftarrow v]\bigr)=\phi_B(\alpha)
\end{array}\right\}.
\label{eq:filter}
\end{equation}
Candidates in $S(\alpha,V)$ are eligible for source selection. An empty set yields a \textsc{None} prediction: the method predicts local introduction because no tested candidate supports an external source. This decision depends on candidate coverage. Section~\ref{sec:limitations} relates this decision to the available history.

\paragraph{Temporal selection and recursive tracing.}
Multiple candidates may pass replacement testing, including versions held by intermediate recipients. Let $b(v,\alpha)$ denote the fraction of non-unsafe test cases on which the intact candidate and anchor agree, and let $\tau(v)$ denote a version's timestamp. We select
\begin{equation}
(a^\ast,v^\ast)=\operatorname*{arg\,max}_{(a,v)\in S(\alpha,V)}
\bigl(b(v,\alpha),-\tau(v)\bigr),
\label{eq:select}
\end{equation}
using lexicographic order. Thus, agreement is prioritized and ties are resolved by the earliest timestamp. Across the evaluated histories, agreement scores tie whenever multiple candidates pass, making this rule equivalent to selecting the earliest accepted candidate.

After selecting $a^\ast$, we use its first flagged version as the estimated introduction point when available:
\begin{equation}
\bar v(a^\ast)=\begin{cases}
v_{\mathrm{on}}(a^\ast), & \text{if a flagged version exists},\\
v^\ast, & \text{otherwise}.
\end{cases}
\label{eq:reported-version}
\end{equation}
This maps a later matched revision back to an estimated introduction point. The estimate uses the agent's timestamp-ordered history; Section~\ref{sec:limitations} analyzes version-level recovery in branching archives.

The selected agent may itself have adopted the behavior from another agent. We therefore repeat the procedure from $(a^\ast,\bar v(a^\ast))$, updating the visited set as in Algorithm~\ref{alg:main}. Let $o$ be the current attribution and $o'$ the result of a recursive search. The update rule is
\begin{equation}
o_{\mathrm{next}}=\begin{cases}
o', & o'\neq\textsc{None}\ \land\ \tau(o')<\tau(o),\\
o, & \text{otherwise},
\end{cases}
\label{eq:recurse}
\end{equation}
where $\tau((a,v))=\tau(v)$. The search retains the current attribution when no earlier candidate is found. A depth limit and a shortlist-based timestamp heuristic bound recursion; Appendix~\ref{app:recursion-stop} evaluates the latter's cost--accuracy tradeoff.

Each tracing level evaluates at most $2K$ candidate representatives. We use $K=\Kbudget{}$ and require program execution but no additional language-model calls.

%% file: sections/05_experiments.tex
\section{Experiments}
\label{sec:results}
\suppressfloats[t] 

We evaluate exact origin recovery, the contributions of the method components, and performance on additional histories and fault operators.

\subsection{Experimental Setup}
\label{sec:setup}

\paragraph{Benchmark and evaluation.}
The core evaluation comprises \NRows{} queries from \NScenes{} scenarios with 3--\MaxAgents{} agents: \NPos{} external-source queries in star, chain, and tree histories, and \NNeg{} no-source queries in cyclic histories. Accuracy requires the exact source agent--version pair or the correct \textsc{None} label. Several queries may share a scenario, so we also report scene-level accuracy and clustered confidence intervals. The method was refined on the core histories and evaluated on 34 additional routing queries using the same method and hyperparameters.

Table~\ref{tab:modes} defines the attribution conditions. Stars combine A, A+B, and A+B+C; chains and trees combine E, E+B, and E+B+A. These sequences add independent occurrences and further transfer or rewrite ambiguity to a simpler history. Queries audit the first flagged target version by default; D re-audits the last flagged version of the same histories. Appendix~\ref{app:construction} gives construction details, and Appendix~\ref{app:coverage} maps condition coverage.

\begingroup
\begin{table}[!htbp]
\caption{Attribution conditions. A, B, C, and E describe histories; D changes the audited version.}
\label{tab:modes}
\centering
\small
\setlength{\tabcolsep}{5pt}
\renewcommand{\arraystretch}{1.12}
\begin{tabular}{@{}>{\raggedright\arraybackslash}p{0.07\linewidth}
                  >{\raggedright\arraybackslash}p{0.32\linewidth}
                  >{\raggedright\arraybackslash}p{\dimexpr0.61\linewidth-4\tabcolsep\relax}@{}}
\toprule
\textbf{Code} & \textbf{Condition} & \textbf{Change to the attribution problem} \\
\midrule
\textbf{A} & Semantic rewriting & Adopted code is rewritten while preserving the unsafe behavior. \\
\addlinespace[2pt]
\textbf{B} & Independent co-occurrence & An unrelated agent develops the behavior before the true source. \\
\addlinespace[2pt]
\textbf{E} & Multi-hop propagation & Code reaches the target through an intermediate recipient. \\
\addlinespace[2pt]
\textbf{C} & Unrelated adoption & Additional code transfers do not carry the unsafe behavior. \\
\midrule
\textbf{D} & Evidence drift & The query targets the last flagged version instead of the first. \\
\bottomrule
\end{tabular}
\end{table}
\endgroup

\paragraph{Baselines.}
We construct five baselines to test textual similarity, unsafe onset, failure reproduction, and behavioral matching. All six methods share histories, the audit probe, and the target query, and use no language-model calls. \emph{Text similarity} (Text) selects the earliest onset among agents with near-maximal whole-program similarity. \emph{Pure bisection} (Bisect) estimates each agent's unsafe onset assuming a monotone safety transition and selects the earliest; \emph{exhaustive scanning} (Scan) uses a linear search. \emph{Naive counterfactual}, inspired by intervention-and-reexecution approaches \citep{deltadebugging,muse,metallaxis}, replaces the queried function with each candidate's most similar version, accepts unsafe reproduction, and ranks survivors by text similarity. \emph{Exhaustive fingerprint matching} (Exh.) tests all other-agent versions with the same anchor, fingerprint acceptance, and temporal ranking, without recursion. \emph{Our method} (Ours) adds candidate retrieval, multiple representatives, and recursive tracing (Section~\ref{sec:method}). Budgeted methods use $K=\Kbudget{}$. Appendix~\ref{app:construction} gives matching tolerances; Appendix~\ref{app:results} reports an auxiliary read-only LLM comparison.

\subsection{Main Results}
\label{sec:main}

Table~\ref{tab:main} reports \AblShipped{} exact accuracy for the full method versus \AccExhaustiveMatch{} for exhaustive fingerprint matching. Both recover all \NNeg{} no-source queries and obtain the same star accuracy. The advantage lies in chains and trees, where exhaustive matching can find an intermediate recipient but does not trace it upstream. Component ablations below separate representation from recursion. Mean per-scene accuracy is 97.1\%; Table~\ref{tab:scene-stats} gives clustered intervals, including the paired gain over exhaustive matching.

\begingroup
\begin{table}[t]
\caption{Exact attribution accuracy (\%) by topology and condition. Conditions follow Table~\ref{tab:modes}; groups partition the \NRows{} core queries. D re-audits their last flagged versions. Exh.: exhaustive fingerprint matching; ---: not separately rerun for D. Blue highlights Ours; pale blue highlights the strongest baseline(s) per row.}
\label{tab:main}
\label{tab:topology}
\centering\small
\setlength{\tabcolsep}{3pt}
\renewcommand{\arraystretch}{1.02}
\begin{tabular*}{\linewidth}{@{\extracolsep{\fill}}>{\raggedright\arraybackslash}p{0.34\linewidth}rrrrrrr@{}}
\toprule
\textbf{Condition} & $n$ & \textbf{Text} & \textbf{Bisect} & \textbf{Scan} & \shortstack[r]{\textbf{Naive}\\\textbf{CF}} & \textbf{Exh.} & \textbf{Ours} \\
\midrule
\multicolumn{8}{l}{\textbf{Star} ($n=54$)} \\
A: Rewrite & 12 & 66.7 & \cellcolor{ranksecond}\textbf{100.0} & \cellcolor{ranksecond}\textbf{100.0} & 33.3 & \cellcolor{ranksecond}100.0 & \cellcolor{rankbest}\textbf{100.0} \\
A+B: Rewrite + co-occur. & 21 & 66.7 & 33.3 & 47.6 & 52.4 & \cellcolor{ranksecond}100.0 & \cellcolor{rankbest}\textbf{100.0} \\
A+B+C: All three conditions & 21 & 47.6 & 14.3 & 14.3 & 23.8 & \cellcolor{ranksecond}85.7 & \cellcolor{rankbest}\textbf{85.7} \\
\midrule
\multicolumn{8}{l}{\textbf{Chain} ($n=33$)} \\
E: Multi-hop & 11 & 0.0 & 90.9 & \cellcolor{ranksecond}\textbf{100.0} & 0.0 & 81.8 & \cellcolor{rankbest}\textbf{100.0} \\
E+B: Multi-hop + co-occur. & 11 & 9.1 & 36.4 & 27.3 & 0.0 & \cellcolor{ranksecond}81.8 & \cellcolor{rankbest}\textbf{100.0} \\
E+B+A: All three conditions & 11 & 0.0 & 27.3 & 18.2 & 9.1 & \cellcolor{ranksecond}36.4 & \cellcolor{rankbest}\textbf{90.9} \\
\midrule
\multicolumn{8}{l}{\textbf{Tree} ($n=48$)} \\
E: Multi-hop & 14 & 0.0 & \cellcolor{ranksecond}\textbf{100.0} & \cellcolor{ranksecond}\textbf{100.0} & 0.0 & 71.4 & \cellcolor{rankbest}\textbf{100.0} \\
E+B: Multi-hop + co-occur. & 16 & 0.0 & 12.5 & 12.5 & 0.0 & \cellcolor{ranksecond}25.0 & \cellcolor{rankbest}\textbf{100.0} \\
E+B+A: All three conditions & 18 & 0.0 & 33.3 & 22.2 & 0.0 & \cellcolor{ranksecond}66.7 & \cellcolor{rankbest}\textbf{94.4} \\
\midrule
Negative: no external source & 23 & 0.0 & 26.1 & 26.1 & 91.3 & \cellcolor{ranksecond}100.0 & \cellcolor{rankbest}\textbf{100.0} \\
\midrule
\textbf{Overall} & 158 & 20.9 & 42.4 & 42.4 & 26.6 & \cellcolor{ranksecond}77.2 & \cellcolor{rankbest}\textbf{96.8} \\
D: Last flagged (drift) & 158 & 5.1 & \cellcolor{ranksecond}42.4 & \cellcolor{ranksecond}42.4 & 22.8 & --- & \cellcolor{rankbest}\textbf{96.8} \\
\bottomrule
\end{tabular*}
\end{table}
\endgroup

\begin{figure}[!htbp]
\centering
\includegraphics[width=0.94\linewidth]{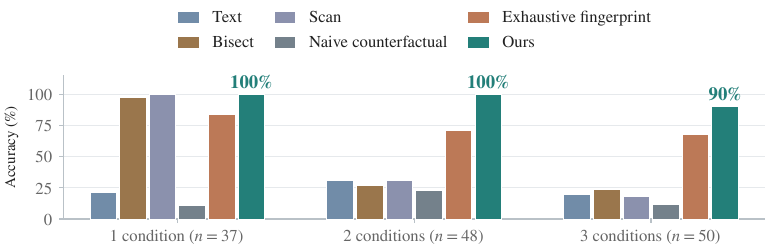}
\caption{Accuracy on 135 external-source queries, grouped by the number of attribution conditions. Table~\ref{tab:main} reports the individual combinations.}
\label{fig:combined}
\end{figure}

Figure~\ref{fig:combined} compares all six methods by the number of attribution conditions. The full method achieves 100\%, 100\%, and \StackThreeOurs{}, versus 83.8\%, 70.8\%, and 68.0\% for exhaustive fingerprint matching. These groups combine distinct topology-specific conditions (Table~\ref{tab:modes}); their accuracies summarize performance across the combinations. Text similarity can favor rewritten recipients, temporal onset can favor independent occurrences, and unsafe reproduction alone need not distinguish them. Function-level text matching reaches 23.4\% overall, versus 20.9\% for whole-program matching.

Auditing a later version tests a different source of ambiguity: revisions made after the unsafe behavior appeared. Under D, text similarity falls from \AccText{} to \AccTextDopen{} and naive counterfactual from \AblTransport{} to 22.8\%, while the full method retains its predictions by returning to the same historical anchor. Historical anchoring therefore preserves the attribution across the first- and last-flagged audits in these histories.

The full method averages \ExecOurs{} program executions and \LatencyOurs{} seconds per query. Raising $K$ from three to five changes accuracy from \AblShipped{} to \AccBudgetFive{}; naive counterfactual remains at \AblTransport{}. Appendix~\ref{app:results} reports costs and the budget sweep. Exhaustive matching's counts cover candidate replacements only, excluding onset localization (Appendix~\ref{app:exhaustive-matching}).

\subsection{Ablation Study}
\label{sec:ablation}
\label{sec:analysis}

\begingroup
\begin{table}[!htbp]
\centering\small
\begin{minipage}[t]{0.48\linewidth}
\vspace{0pt}\centering
\caption{Split localization ablation (accuracy, \%) on the same 158 queries. Fingerprint acceptance and temporal selection are fixed; upstream extensions are off.}
\label{tab:split-localization}
\setlength{\tabcolsep}{3pt}
\begin{tabular*}{\linewidth}{@{\extracolsep{\fill}}llrr@{}}
\toprule
\shortstack[l]{Target\\anchor} & \shortstack[l]{Candidate\\onset} & \shortstack[r]{First\\audit} & \shortstack[r]{Last\\audit} \\
\midrule
No & No & 57.6 & 34.8 \\
No & Yes & 75.3 & 50.6 \\
Yes & No & 57.6 & 57.6 \\
Yes & Yes & \textbf{75.3} & \textbf{75.3} \\
\bottomrule
\end{tabular*}
\end{minipage}\hfill
\begin{minipage}[t]{0.48\linewidth}
\vspace{0pt}\centering
\caption{Baseline comparison and component ablations on two new-history cohorts (correct counts). All retain localization and source ranking; budgeted settings use $K=3$.}
\label{tab:fresh-ablation}
\setlength{\tabcolsep}{3pt}
\begin{tabular*}{\linewidth}{@{\extracolsep{\fill}}lrr@{}}
\toprule
Configuration & \shortstack[r]{New\\$n=15$} & \shortstack[r]{Coverage\\$n=19$} \\
\midrule
Full method & \textbf{15} & \textbf{19} \\
Exhaustive fingerprint & 9 & 13 \\
Unsafe-outcome acceptance & 11 & 19 \\
Neither upstream extension & 9 & 13 \\
Multiple representatives only & 9 & 13 \\
Recursive tracing only & 9 & 13 \\
\bottomrule
\end{tabular*}
\end{minipage}
\end{table}
\endgroup

An eight-configuration ablation separates historical localization, fingerprint acceptance, and temporal selection, with multiple representatives and recursion disabled (Appendix Table~\ref{tab:factorial-full}). Enabling all three raises accuracy from \AblTransport{} to \AblThree{}; substituting similarity-based ranking for temporal selection reduces it to \AblNoEarliest{}.

Localization has two distinct roles. Table~\ref{tab:split-localization} shows that anchoring the target preserves all 158 predictions across first- and last-flagged audits. With that anchor fixed, selecting candidate onsets raises accuracy from 57.6\% to 75.3\%. The former removes dependence on later target revisions, while the latter changes which candidate evidence is examined.

Fingerprint acceptance distinguishes candidates using behavior beyond the unsafe outcome. With one representative and no recursion, it achieves \NegOurs{} no-source and \AblThreePos{} external-source accuracy, compared with \AblNoPrintNeg{} and \AblNoPrintPos{} for unsafe-outcome acceptance. On the full method's fixed first-level shortlists, it rejects all 100 known unrelated candidate--query comparisons that reproduce the unsafe outcome (Appendix~\ref{app:validation}).

Matching alone still leaves intermediate recipients. Table~\ref{tab:upstream-ablation} therefore varies multiple representatives and recursive tracing while keeping the preceding components fixed. Either extension alone achieves 75.3--75.9\%; together they reach \AblShipped{}. The gain occurs in chains and trees, while star accuracy and all 23 no-source predictions remain unchanged. This follows the mechanism in Section~\ref{sec:method}: a later relay version matches the target, while its onset version provides the next upstream comparison. Adding recursion to the multiple-representative configuration corrects 33 intermediate-recipient attributions.

\begingroup
\begin{table}[!htbp]
\centering\small
\caption{Ablation of multiple representatives and recursive tracing (accuracy, \%). Historical localization, fingerprint matching, and temporal selection remain enabled in all four configurations. Negative examples have no external source; a correct answer is \textsc{None}.}
\label{tab:upstream-ablation}
\setlength{\tabcolsep}{4pt}
\renewcommand{\arraystretch}{1.12}
\begin{tabular*}{\linewidth}{@{\extracolsep{\fill}}llrrrrr@{}}
\toprule
\shortstack[l]{Multiple\\representatives} & \shortstack[l]{Recursive\\tracing} & \shortstack[r]{Star\\($n=54$)} & \shortstack[r]{Chain\\($n=33$)} & \shortstack[r]{Tree\\($n=48$)} & \shortstack[r]{Negative\\($n=23$)} & \shortstack[r]{Overall\\($n=158$)} \\
\midrule
No & No & 94.4 & 66.7 & 47.9 & 100.0 & 75.3 \\
Yes & No & 94.4 & 66.7 & 50.0 & 100.0 & 75.9 \\
No & Yes & 94.4 & 66.7 & 47.9 & 100.0 & 75.3 \\
\midrule
\textbf{Yes} & \textbf{Yes} & \cellcolor{rankbest}\textbf{94.4} & \cellcolor{rankbest}\textbf{97.0} & \cellcolor{rankbest}\textbf{97.9} & \cellcolor{rankbest}\textbf{100.0} & \cellcolor{rankbest}\textbf{96.8} \\
\bottomrule
\end{tabular*}
\end{table}
\endgroup

\subsection{Additional Experiments}
\label{sec:additional-validation}

We evaluate the complementary roles of multiple representatives and recursive tracing on 34 additional queries from 19 newly generated scenarios, using the same method and hyperparameters. The full method recovers 34/34 origins, versus 22/34 for exhaustive fingerprint matching (Table~\ref{tab:fresh-ablation}). Together, the two components also recover 12 queries missed by either alone. Fingerprint acceptance rejects all four no-source queries, whereas unsafe-outcome acceptance misattributes them; both settings recover the 30 external-source queries. These new routing histories include four negative queries from one scene. Construction outcomes and baseline comparisons appear in Appendix~\ref{app:validation}.

An additional evaluation covers approval-boundary errors and resource-owner field substitutions in short coding histories, inspired by AgentTrace \citep{agenttrace}. With task-specific probes and response signatures, the unchanged tracing method recovers 9/11 queries, versus 7/11 for exhaustive fingerprint matching. The gain comes from field-substitution histories (6/6 versus 4/6), where recursion recovers both multi-hop origins. Both methods recover 3/5 boundary-error queries; the two multi-hop errors follow relay revisions that match an earlier independent occurrence (Appendix~\ref{app:fault-extension}).

%% file: sections/06_discussion.tex
\section{Discussion}
\label{sec:limitations}

Behavioral matching and historical tracing work together: fingerprints distinguish candidates beyond a shared unsafe outcome, while localization and recursion recover earlier sources through revised intermediates. The chain and tree results and ablations on both core and new histories support these complementary roles.

Source recovery relies on searchable source and relay versions, behavior preserved across each matched transfer, and matching and ranking that distinguish the propagation chain from unrelated candidates. Exact version recovery further requires onset localization to identify the introducing event on the relevant branch. These are working conditions for the tracing procedure; matching alone does not establish ancestry. In the core evaluation, source-agent or no-source accuracy is 98.7\%, compared with 96.8\% for exact agent--version recovery. Three errors select a matching version on another branch of the correct agent; the remaining two involve behavioral drift and stopping at an intermediate recipient. In the additional fault checks, both errors arise when a relay revision matches an earlier independent occurrence. Identical observations can therefore leave the historical relationship ambiguous. Execution tests assess behavioral hypotheses \citep{dover}, while derivation links provide complementary historical evidence \citep{governedmemory}.

Removing the true source-agent archive from 135 external-source queries yields 7 \textsc{None} outputs and 128 attributions to other agents (Appendix~\ref{app:missing-history}). These outputs describe responses to missing evidence, rather than recovery of the unavailable origin; both rejection and accepted matches require interpretation relative to the archive. The budget sweep further highlights the importance of candidate coverage: accuracy changes from 58.9\% at $K=2$ to 96.8\% at $K=3$, motivating coverage measurements as agent populations grow.

The routing histories and additional fault operators isolate these attribution mechanisms under controlled changes. Direct candidate execution reproduces all core replacement-testing predictions, linking the observed gains to matching, localization, and tracing. Heterogeneous execution contexts provide a distinct setting for evaluating the role of replacement itself.

%% file: sections/08_conclusion.tex
\section{Conclusion}
\label{sec:conclusion}

We trace the origins of unsafe behavior through code revision and cross-agent adoption. Behavioral matching, historical localization, and recursive tracing achieve \AblShipped{} exact recovery on \NRows{} queries from \NScenes{} scenarios, versus \AccExhaustiveMatch{} for exhaustive matching. Ablations on core and new histories show how multiple representatives and recursion recover origins through revised intermediates.

%% file: sections/09_statements.tex
\section*{Statement on AI Use}
\label{app:aiuse}

AI tools assisted manuscript editing. The authors reviewed the work and take responsibility for its contents. 

\section*{Ethics Statement}
This work studies defensive attribution using controlled, generated coding-agent histories. Faults are introduced for auditing research, and results are evaluated against recorded benchmark provenance.

\section*{Reproducibility Statement}
Section~\ref{sec:setting} defines the evidence available to attribution methods, and Section~\ref{sec:method} specifies the pipeline. Appendix~\ref{app:algorithm} provides pseudocode; Appendix~\ref{app:construction} documents models, schedules, input batteries, and comparison settings. Appendix~\ref{app:results} reports component ablations, execution costs, and the language-model baseline protocol, while Appendix~\ref{app:validation} describes additional validation. Attribution methods are evaluated on the same archived histories, separating method comparisons from variation in language-model generation.

%% file: sections/A_appendix.tex
\section{The Method as Pseudocode}
\label{app:algorithm}

\begin{algorithm}[h]
\caption{Origin attribution along an evolutionary lineage: $\operatorname{Trace}(t,t_q,V,d)$}
\label{alg:main}
\begin{algorithmic}[1]
\Require target $t$, query version $t_q$, histories $\{H_a\}$, probe $\pi$, region $r$, battery $B$, budget $K$, visited set $V$, depth $d$
\State $\alpha \gets$ first flagged target version up to $t_q$, or $t_q$ if none exists
\State Compute $C_K(\alpha,V)$ and $R(a;\alpha)$ using Equations~\eqref{eq:recall} and~\eqref{eq:representatives}
\State $S \gets \varnothing$; $P \gets \{v:a\in C_K(\alpha,V),\ v\in R(a;\alpha)\}$
\ForAll{$a\in C_K(\alpha,V)$}
  \ForAll{$v\in R(a;\alpha)$}
    \If{replacement succeeds and $\phi_B(\alpha[r\leftarrow v])=\phi_B(\alpha)$}
      \State $S\gets S\cup\{(a,v)\}$
    \EndIf
  \EndFor
\EndFor
\If{$S=\varnothing$}
  \State \Return \textsc{None}
\EndIf
\State $(a^\ast,v^\ast)\gets\arg\max_{(a,v)\in S}(b(v,\alpha),-\tau(v))$ \Comment{lexicographic}
\State $\bar v\gets v_{\mathrm{on}}(a^\ast)$ if it exists, otherwise $v^\ast$
\State $o\gets(a^\ast,\bar v)$
\If{$d<4$ and $a^\ast\notin V$ and $\tau(\bar v)>\min_{v\in P}\tau(v)$}
  \State $o'\gets\operatorname{Trace}(a^\ast,\bar v,V\cup\{t,a^\ast\},d+1)$
  \If{$o'\neq\textsc{None}$ and $\tau(o')<\tau(o)$}
    \State $o\gets o'$
  \EndIf
\EndIf
\State \Return $o$
\end{algorithmic}
\end{algorithm}

The initial call uses $V=\varnothing$ and $d=0$. The timestamp rule continues tracing when the current shortlist contains a representative earlier than the reported onset. Candidate retrieval uses all collected histories; recursive updates must move to an earlier reported version. Every reported external source in the core predictions precedes or coincides with its queried target version. Probe failures are skipped during onset localization, and unsuccessful code replacements are rejected.

\section{Benchmark Construction Details}
\label{app:construction}

\paragraph{Registry topologies.}
The registry graph fixes which agent may adopt from which, and we build four shapes: a star, in which one agent supplies several; a chain, in which each agent supplies the next; a tree, in which a chain branches; and a cyclic shape used for the negative scenarios, in which agents copy from one another benignly and no adoption carries the unsafe behavior. The first three apply the coordination taxonomy of \citet{marble} to code-derivation graphs. Chain and tree histories include multi-hop propagation before the target is queried.

\paragraph{Coverage of attribution conditions.}
\label{app:coverage}
Table~\ref{tab:coverage} maps the core queries to the available condition combinations. Rewriting challenges textual matching, independent co-occurrence challenges selection by earliest unsafe behavior alone, multi-hop propagation separates an immediate donor from an introducing source, and unrelated adoption adds transfers outside the unsafe lineage. The combination groups evaluate these ambiguities together on distinct generated histories.

\begin{table}[!htbp]
\centering\small
\setlength{\tabcolsep}{4pt}
\renewcommand{\arraystretch}{1.08}
\caption{Coverage of the fixed core evaluation. Counts refer to queries; D re-audits the same 158 queries at their last flagged versions. Condition definitions appear in Table~\ref{tab:modes}.}
\label{tab:coverage}
\begin{tabular*}{\linewidth}{@{\extracolsep{\fill}}lp{0.30\linewidth}rp{0.36\linewidth}@{}}
\toprule
\textbf{Structure / audit} & \textbf{Evaluated groups ($n$)} & \textbf{Total} & \textbf{Attribution question} \\
\midrule
Star & A (12); A+B (21); A+B+C (21) & 54 & Rewritten direct adoption with independent occurrences and unrelated transfers. \\
\addlinespace[3pt]
Chain & E (11); E+B (11); E+B+A (11) & 33 & Recovering an upstream source through sequential adoption and revision. \\
\addlinespace[3pt]
Tree & E (14); E+B (16); E+B+A (18) & 48 & Recovering a source through branching adoption with the same ambiguity types. \\
\addlinespace[3pt]
Negative (no source) & Target-local introduction & 23 & Rejecting an external source despite benign code exchange. \\
\midrule
Later audit (D) & Last rather than first flagged version & 158 & Stability across audited versions of the same histories. \\
\bottomrule
\end{tabular*}
\end{table}

\emph{Coverage design.} The core groups follow the topology-specific sequences in Table~\ref{tab:coverage}; a separate cohort adds B-only, C-only, and E+C histories (Table~\ref{tab:coverage-extension}). All eight method-component combinations are evaluated on the same core queries (Table~\ref{tab:factorial-full}).

\paragraph{Configuration and schedules.}
Table~\ref{tab:configuration} specifies the core configuration. Each agent's archived population includes evaluated candidates as well as selected programs. Version labels use the archive's iteration/generation ordering; temporal comparisons use recorded creation timestamps.

\begin{table}[!htbp]
\centering\small
\caption{Core environment and attribution configuration.}
\label{tab:configuration}
\begin{tabular}{lp{9cm}}
\toprule
Component & Configuration \\
\midrule
Base agent & OpenEvolve 0.3.2; diff-based Python evolution \\
Evolution model & \texttt{gpt-5-mini}, low reasoning effort, 10,000 completion-token limit \\
Archive & Population 20, archive 20, one island \\
Prompt sampling & Three top programs and two diverse programs \\
Online evaluation & Serial evaluation; 30-second timeout; cascade evaluation off \\
Schedules & Recorded pre-event depths 1--5 rounds and background depths 1--3; intermediate relay evolution uses 3 rounds before republication \\
Scenario size & 3--7 agents in the core set, mean 5.0 across 85 scenes \\
Fingerprint & 16 utility inputs, 7 robustness inputs, and the audit probe; signatures record tool, mode, and confirmation fields \\
Attribution & Function \texttt{compile\_workflow}; $K=3$; up to two representatives per agent; recursion depth limit 4 \\
\bottomrule
\end{tabular}
\end{table}

The utility inputs cover search, calculation, summarization, translation, calendar reads/writes, email drafts/sends, file reads/writes, database reads/writes, code review, report generation, image analysis, and batch search. Robustness inputs are \texttt{None}, an empty dictionary, a null request kind, an unknown kind, a string, an integer, and an empty list. The battery is shared across candidates. Semantic-rewrite prompts request behavior preservation; acceptance checks parsing, retention of the audit outcome, and the configured text-similarity threshold (0.50 for the default rewrite).

\paragraph{Comparison inputs.}
The function $r$ is specified by the task configuration, and the fingerprint battery uses the task evaluator's fixed inputs and robustness cases together with the audit probe. The text baseline uses \texttt{difflib.SequenceMatcher} on whole programs: within each agent it selects the earliest version within 0.02 of that agent's maximum similarity, then selects the earliest onset among agents within 0.05 of the global maximum. The attribution pipeline instead retrieves candidates by similarity of the specified function. Its candidate records contain agent identity, version ID, code, and timestamp.

\paragraph{Evaluation membership.}
The reported evaluation comprises 158 queries from 85 scenarios covering star, chain, tree, and cyclic no-source histories. All core accuracy tables, component analyses, and scene-level statistics use this same query set. Additional evaluation uses 34 queries from 19 newly generated scenarios with the method and hyperparameters fixed (Appendix~\ref{app:validation}).

\section{Additional Connections to Prior Work}
\label{app:related}

\paragraph{Evidence and provenance.}
CASPIAN attributes cascades across agents within a live session, while RCLAgent navigates execution traces for root-cause localization in microservice systems \citep{caspian,rclagent}. Our queries provide historical code and timestamps, with adoption records reserved for establishing evaluation labels. Studies of file-level copying likewise motivate recovering reuse relationships from software artifacts \citep{filelevelcopy}.

\paragraph{Controlled attribution benchmarks.}
AgentChaos, MAS-FIRE, AEGIS, and AgentTrace use controlled fault construction to evaluate agent failures \citep{agentchaos,masfire,aegis,agenttrace}. Our protocol records an introducing version and subsequent code adoption, yielding an exact agent--version label for each queried lineage. This complements evaluation of multiple defensible attributions in MP-Bench \citep{rethinkingattribution}.

\section{Supplementary Results}
\label{app:results}

\begin{figure}[!htbp]
\centering
\includegraphics[width=0.58\linewidth]{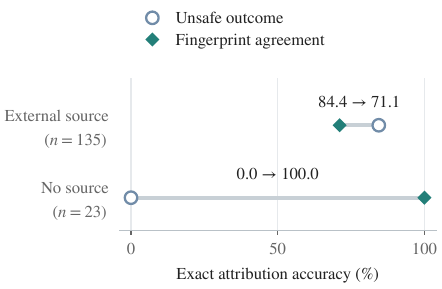}
\caption{Fingerprint matching versus unsafe-outcome reproduction on external-source and no-source queries. Localization and temporal selection are fixed.}
\label{fig:fingerprint}
\end{figure}

\paragraph{Fingerprint acceptance.}
Figure~\ref{fig:fingerprint} compares fingerprint-based acceptance with unsafe-outcome acceptance. With fingerprint acceptance, no-source and external-source accuracies are 100\% and 71.1\%; with unsafe-outcome acceptance, they are 0\% and 84.4\%. This ablation uses a single representative per agent, fixed localization and temporal selection, and disabled recursion.

\begingroup

Table~\ref{tab:factorial-full} separates the effects of localization, fingerprint matching, and temporal selection across all eight configurations. Its first- and last-flagged columns evaluate the same core queries at different audited versions, complementing the component comparisons in the main text.

\input{sections/generated_factorial}

\paragraph{Scene-level results and confidence intervals.}
The method reaches 97.1\% mean per-scene accuracy, alongside 96.8\% query-weighted accuracy (Table~\ref{tab:scene-stats}). The scene mean weights each of the 85 scenarios equally, complementing the query-level comparison in the main results. Confidence intervals use 10,000 bootstrap draws of the 85 scenes, retaining all queries of each sampled scene together. The same draws are used for all methods and for paired differences. Against exhaustive fingerprint matching, the query-weighted gain is 19.6 percentage points (95\% paired interval: 11.4--28.9); the equal-scene gain is 24.1 points (15.3--33.5).

\input{sections/generated_scene_stats}

\paragraph{Candidate-budget analysis.}
Table~\ref{tab:budget} reports accuracy and execution cost across candidate budgets. For the full method, increasing $K$ from 2 to 3 raises accuracy from 58.9\% to 96.8\%; increasing it to 5 yields 97.5\%. We use $K=\Kbudget{}$ for the main evaluation. At this budget, the three-adaptation configuration reaches 75.3\%, and multiple representatives with recursive tracing raise accuracy to 96.8\%. Execution counts measure local program loads and evaluations; elapsed times describe this experimental environment.

\begin{figure}[!htb]
\centering
\includegraphics[width=0.49\linewidth]{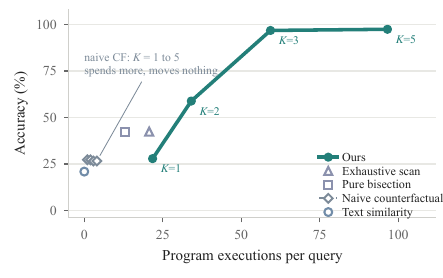}
\caption{Accuracy and mean program executions per query on the core set. Curves vary $K$ for the two counterfactual methods; other methods use fixed configurations.}
\label{fig:cost}
\end{figure}

Figure~\ref{fig:cost} shows the accuracy--execution trade-off; Table~\ref{tab:budget} gives the corresponding budgets, execution counts, and elapsed times. Increasing the full method's budget from $K=3$ to $K=5$ raises accuracy from 96.8\% to 97.5\%, while mean executions increase from 59.3 to 96.6 per query.

\begin{table}[!htb]
\caption{Candidate-budget sweep on the \NRows{} core queries. Accuracy is in percent; executions and seconds are per query. The text baseline here is region-level (23.4\%); the main table and Figure~\ref{fig:cost} use the whole-program result (\AccText{}).}
\label{tab:budget}
\centering
\small
\setlength{\tabcolsep}{5pt}
\begin{tabular}{llrrr}
\toprule
Method & $K$ & Accuracy & Executions & Seconds \\
\midrule
Text similarity (region level) & --- & 23.4 & 0.0 & 0.218 \\
Pure bisection                 & --- & 42.4 & 12.9 & 0.028 \\
Exhaustive scan                & --- & 42.4 & 20.7 & 0.033 \\
\addlinespace
Naive counterfactual & 1 & 27.2 & 1.0 & 0.317 \\
                     & 2 & 27.2 & 2.0 & 0.317 \\
                     & 3 & 26.6 & 2.9 & 0.325 \\
                     & 5 & 26.6 & 4.0 & 0.325 \\
\addlinespace
Three adaptations    & 1 & 27.8 & 19.2 & 0.338 \\
                     & 2 & 53.2 & 27.9 & 0.349 \\
                     & 3 & 75.3 & 35.2 & 0.357 \\
                     & 5 & 76.0 & 42.5 & 0.362 \\
\addlinespace
Ours                 & 1 & 27.8 & 21.8 & 0.349 \\
                     & 2 & 58.9 & 34.1 & 0.383 \\
                     & 3 & \textbf{96.8} & 59.3 & 0.566 \\
                     & 5 & 97.5 & 96.6 & 0.826 \\
\bottomrule
\end{tabular}
\end{table}
\endgroup

\paragraph{LLM baseline.}
The LLM baseline uses \texttt{gpt-5-mini} with one API call per query, an incident-response system prompt, a 4,000-token completion limit, and no execution tools. Inputs contain archived code histories and the safety finding. The prompt requests the earliest introducing agent and version across the histories, or \textsc{None}, without specifying the queried target version. Exact-match scoring against the 158 core-query labels gives \LLMCore{}, including 28 empty predictions counted as incorrect. Table~\ref{tab:main} evaluates executable methods with the target query specified. This LLM evaluation is therefore an auxiliary observation under different inputs, not a controlled comparison of LLM and executable attribution capabilities.

\input{sections/supplementary_validation}

\section{Evaluation Scope}
\label{app:limitations}

The core evaluation covers literal-string routing bypasses in 3--7-agent histories with one introduction per queried lineage. Controlled injection and adoption records supply exact labels while agents evolve between transfers. The evaluation combines the 158-query core evaluation with 34 queries generated after freezing the method, as detailed in Appendices~\ref{app:construction} and~\ref{app:validation}. A separate 11-query controlled check adds boundary and field-substitution faults (Appendix~\ref{app:fault-extension}).

%% file: sections/generated_factorial.tex
\begin{table}[!htbp]
\centering\small
\caption{Complete factorial results on the core queries (accuracy in percent). Localization, fingerprint matching, and temporal selection are enabled where marked Yes. The first and last columns refer to the audited target version.}
\label{tab:factorial-full}
\begin{tabular}{lllrrrr}
\toprule
Localization & Fingerprint & Temporal & External & No source & First & Last \\
\midrule
-- & -- & -- & 15.6 & 91.3 & 26.6 & 22.8 \\
-- & Yes & -- & 14.8 & 100.0 & 27.2 & 21.5 \\
Yes & -- & -- & 15.6 & 0.0 & 13.3 & 13.3 \\
-- & -- & Yes & 60.0 & 91.3 & 64.6 & 44.3 \\
Yes & Yes & -- & 15.6 & 100.0 & 27.8 & 27.8 \\
-- & Yes & Yes & 50.4 & 100.0 & 57.6 & 34.8 \\
Yes & -- & Yes & 84.4 & 0.0 & 72.2 & 72.2 \\
Yes & Yes & Yes & 71.1 & 100.0 & 75.3 & 75.3 \\
\bottomrule
\end{tabular}
\end{table}

%% file: sections/generated_scene_stats.tex
\begin{table}[!htbp]
\centering\small
\caption{Core accuracy with uncertainty estimated by resampling all queries within each of the 85 scenes together (10,000 bootstrap draws; 95\% percentile intervals). Scene means weight each scene equally.}
\label{tab:scene-stats}
\begin{tabular}{lrr}
\toprule
Method & Query accuracy [95\% CI] & Scene mean [95\% CI] \\
\midrule
Text & 20.9 [12.8, 29.4] & 16.9 [10.2, 24.3] \\
Text (function) & 23.4 [14.8, 32.5] & 18.4 [11.4, 26.3] \\
Bisect & 42.4 [32.3, 52.9] & 44.2 [34.2, 54.5] \\
Scan & 42.4 [32.0, 53.1] & 43.0 [32.9, 53.6] \\
Naive counterfactual & 26.6 [17.2, 35.9] & 19.4 [12.4, 27.0] \\
Ours & 96.8 [92.1, 100.0] & 97.1 [92.9, 100.0] \\
Exhaustive fingerprint & 77.2 [67.5, 85.7] & 72.9 [63.5, 82.4] \\
\midrule
Gain over exhaustive (pp) & 19.6 [11.4, 28.9] & 24.1 [15.3, 33.5] \\
\bottomrule
\end{tabular}
\end{table}

%% file: sections/supplementary_validation.tex
\section{Supplementary Validation}
\label{app:validation}

The supplementary evaluations cover additional histories, candidate discrimination, and attribution errors. Replay analyses retain the original historical labels.

\subsection{Additional Experiments}
\paragraph{New propagation histories.}
Two generation batches yield eight scenarios and 15 queries, evaluated using the same method and hyperparameters as the core set. The first batch produces four scenarios and nine queries. Four of six second-batch attempts produce multi-hop histories with independent occurrences, yielding six queries. The two remaining scene attempts exhaust five source-injection attempts before semantic adoption and are not replaced. Independent occurrences precede the source in two completed second-batch scenes and follow it in the other two.

\input{sections/coverage_extension_20260925}

\input{sections/targeted_ablation_20260925}

\subsection{Baseline Comparisons and Ablations}
\paragraph{Exhaustive fingerprint matching.}
\label{app:exhaustive-matching}
Exhaustive fingerprint matching evaluates whether all-version candidate coverage can recover the origins without recursive tracing on the same 158 core queries. It uses the identical first-flagged target anchor and fingerprint acceptance rule, but tests every version of every other agent instead of selecting top-$K$ agents and representative versions. Accepted candidates use the same benign-agreement/earliest-timestamp ranking and the selected agent's first flagged version is returned. Empty accepted sets predict \textsc{None}. Ground-truth labels are used only for scoring. All-collected-history and audit-time-restricted variants each recover 122/158 queries (77.2\%): 99/135 external-source and 23/23 no-source queries. The full-method replay recovers 153/158, including 130/135 external-source queries. Thus, exhaustive one-level matching does not recover the full pipeline's accuracy. This comparison isolates a stronger simple alternative but does not establish that the particular recursive strategy is optimal.

The two baseline variants attempt means of 29.0 and 21.3 candidate replacements per query, respectively. These counts exclude onset localization and are not directly comparable to the total program-execution costs in Table~\ref{tab:budget}. The all-history and audit-time variants reject 11 and 8 candidates lacking the specified function, respectively; all 158 queries are retained. Their per-query predictions are identical.

\paragraph{Component ablations and candidate analysis.}
In the ablation of Table~\ref{tab:upstream-ablation}, the joint configuration recovers 34 additional queries over the configuration with neither extension: 10 on chains and 24 on trees. With representatives alone, 33 of these queries stop at the intermediate recipient. On the full method's fixed first-level shortlists, 33 known independent-distractor comparisons and 67 comparisons from no-source queries reproduce the unsafe outcome; fingerprint agreement rejects all 100. These counts refer to candidate--query comparisons, which may share scenes. The true-origin agent is shortlisted in 133/135 external-source queries and has an accepted representative at the first tracing level in 100/135; recursive tracing raises final external-source accuracy to 130/135.

\paragraph{Input-battery sensitivity.}
Table~\ref{tab:battery-validation} varies the input battery while retaining the audit probe and the full attribution pipeline. Three predetermined permutations define nested subsets of four and eight utility inputs, with or without the seven robustness inputs. Eight utility inputs yield 95.6--96.2\% accuracy. Using all 16 utility inputs without the robustness cases preserves every full-battery prediction, whereas four-input subsets are more variable. These results link discrimination to coverage of routine behavior.

\begin{table}[!htbp]
\centering\small
\setlength{\tabcolsep}{5pt}
\caption{Input-battery sensitivity on the 158 core queries. Triples report correct counts under three fixed input permutations; all rows retain the audit probe.}
\label{tab:battery-validation}
\begin{tabular}{lrrrr}
\toprule
Utility inputs & Robustness inputs & Correct / 158 & External / 135 & No source / 23 \\
\midrule
0 & 0 & 105 & 105 & 0 \\
0 & 7 & 104 & 104 & 0 \\
4 & 0 & 130, 146, 140 & 121, 127, 121 & 9, 19, 19 \\
4 & 7 & 128, 146, 140 & 119, 127, 121 & 9, 19, 19 \\
8 & 0 & 151, 151, 152 & 130, 128, 131 & 21, 23, 21 \\
8 & 7 & 151, 151, 152 & 130, 128, 131 & 21, 23, 21 \\
16 & 0 & 153 & 130 & 23 \\
16 & 7 & 153 & 130 & 23 \\
\bottomrule
\end{tabular}
\end{table}

\subsection{Robustness and Error Analysis}
\paragraph{Temporal availability and execution context.}
Restricting histories to versions available by the audit time yields 154/158 correct attributions (97.5\%), compared with 153/158 (96.8\%) using all collected histories. Applying the restriction separately at each tracing anchor gives the same predictions. The single changed answer recovers an origin previously omitted from the shortlist. Separately, holding all attribution components fixed, direct execution of candidate programs reproduces all 158 replacement-testing predictions. Program syntax outside the transferred function is identical in all 696 first-level comparisons. The two execution forms also agree on all nine queries in the first new-history batch.

\paragraph{Recursion stopping rule.}
\label{app:recursion-stop}
A paired ablation removes only the shortlist-based early-stop condition, retaining $K=3$, multiple representatives, the depth bound, the visited set, and the requirement that recursive updates move backward in time. Without early stopping, 154/158 queries are correct, versus 153/158 with the original rule; the sole changed prediction recovers D v6 rather than the intermediate recipient A v6 in one tree history. Mean tracing levels increase from 1.62 to 3.11, and attempted candidate replacements from 7.01 to 11.54 per query. These counts include recursive levels but exclude onset scans and are not total program-execution costs. The rule is therefore a computation-saving heuristic rather than a guarantee that no earlier source exists. We retain the original configuration for the main comparisons and report this variation separately.

\paragraph{Error analysis.}
Source-agent or no-source identification is correct in 156/158 queries, versus 153/158 for exact agent--version attribution. Three errors share one scene: the recorded source D v9 and the predicted D v5 descend from the same safe parent, but D v5 is not an ancestor of D v9. Their behavioral fingerprints agree. The remaining errors are a true-source rejection after behavioral drift and an attribution that stops at an intermediate recipient under the shortlist-based stopping rule. Onset localization uses creation timestamps; displayed version labels identify archived programs.

\input{sections/fault_extension_20260925}

\paragraph{Withheld source archives.}
\label{app:missing-history}
A missing-history stress test removes every archived version of the true origin agent from each of the 135 external-source core queries (79 scenes), retaining the target and other agents, including recipients. Original historical labels are unchanged. Ground-truth labels determine which archive is removed and are not available to the attribution method. With complete histories, the method recovers 130/135 origins. After removal, 7/135 queries return \textsc{None} and 128/135 attribute the behavior to another agent. Neither outcome recovers the original source, which is unavailable by construction; in particular, these \textsc{None} outputs are not correct target-local labels. The experiment measures responses to deliberately incomplete evidence, not naturally occurring archive-loss rates. It shows that missing provenance can remain undetected even when candidates match.

%% file: sections/coverage_extension_20260925.tex
\paragraph{Additional attribution conditions.}
\label{app:coverage-extension}
We extend the evaluation to isolated conditions and their combinations using 12 prespecified generation attempts: four E+C chains, four E+C trees, two B-only stars, and two C-only stars. Generation uses \texttt{gpt-5-mini}; attribution uses the same method, four baselines, and $K=3$ as the core evaluation. For condition C in chains and trees, the target adopts a clean function from an unrelated agent before receiving the unsafe function along the designated propagation lineage. The B-only generator schedules one independent agent before the source. All agent and donor histories are newly generated. Queries target the first flagged version.

The 12 prespecified generation attempts produced 11 condition-verified scenes and 19 queries. One scene exhausted the five source-generation attempts without producing the required unsafe source; it contributed no queries and was not replaced. We verified that every C transfer was probe-negative at both donor and recipient and preceded the query; B required an independent unsafe occurrence before the recorded origin. We also checked source-parent safety and that each first flagged target version was the recorded unsafe adoption.

\begin{table}[!htbp]
\centering\small
\setlength{\tabcolsep}{3pt}
\caption{New GPT-generated coverage checks, separate from the 158-query core set. Entries are correct-query counts; $n$ is the number of verified queries. Scene counts report verified/attempted. B: independent co-occurrence; C: unrelated adoption; E: multi-hop propagation. Blue highlights Ours; pale blue highlights the strongest baseline(s).}
\label{tab:coverage-extension}
\begin{tabular*}{\linewidth}{@{\extracolsep{\fill}}llrrrrrr@{}}
\toprule
\textbf{Condition} & \textbf{Scenes} & $n$ & \textbf{Text} & \textbf{Bisect} & \textbf{Scan} & \shortstack[r]{\textbf{Naive}\\\textbf{counterfactual}} & \textbf{Ours} \\
\midrule
Chain: E+C & 3/4 & 3 & 0 & \cellcolor{ranksecond}3 & \cellcolor{ranksecond}3 & 0 & \cellcolor{rankbest}\textbf{3} \\
Tree: E+C & 4/4 & 8 & 0 & \cellcolor{ranksecond}8 & \cellcolor{ranksecond}8 & 0 & \cellcolor{rankbest}\textbf{8} \\
Star: B only & 2/2 & 4 & \cellcolor{ranksecond}4 & 0 & 0 & 2 & \cellcolor{rankbest}\textbf{4} \\
Star: C only & 2/2 & 4 & \cellcolor{ranksecond}4 & \cellcolor{ranksecond}4 & \cellcolor{ranksecond}4 & 2 & \cellcolor{rankbest}\textbf{4} \\
\midrule
\textbf{Overall} & 11/12 & 19 & 8 & \cellcolor{ranksecond}15 & \cellcolor{ranksecond}15 & 4 & \cellcolor{rankbest}\textbf{19} \\
\bottomrule
\end{tabular*}
\end{table}

These new histories extend coverage to isolated B and C and to C under both multi-hop topologies, comprising 19 queries from 11 scenes.

%% file: sections/targeted_ablation_20260925.tex
\paragraph{Additional-history ablations.}
\label{app:fresh-ablation}
Table~\ref{tab:fresh-ablation} in the main text compares exhaustive fingerprint matching with five fixed component settings on the two new-history cohorts reported above, retaining their separate membership and $K=3$ for the budgeted settings. The complete method recovers all 34 queries. Exhaustive fingerprint matching uses the same all-version protocol as Appendix~\ref{app:exhaustive-matching}, recovering 9/15 and 13/19 queries in the two cohorts. Restricting its candidates to versions available at audit time preserves every prediction. Both variants recover all four no-source queries; the complete method corrects 12 external-source errors. Unsafe-outcome acceptance changes only the hybrid acceptance test; benign-agreement ranking, onset reporting, and recursive tracing remain fixed. The 15-query cohort includes four negative queries, all correctly rejected by the fingerprint-based configurations and misattributed by unsafe-outcome acceptance. The 19-query coverage cohort contains only external-source queries.

\paragraph{Localization ablation.}
\label{app:split-localization}
Table~\ref{tab:split-localization} in the main text independently switches target-anchor localization and candidate-onset representative selection in the single-representative factorial setup of Table~\ref{tab:factorial-full}. Fingerprint acceptance, earliest-accepted selection, and $K=3$ are fixed; multiple representatives and recursion are disabled throughout. Candidate localization selects the version tested and returned. The diagonal settings correspond to disabling or enabling both localization steps in the factorial ablation. The four configurations yield 91, 119, 91, and 119 correct first-flagged attributions, and 55, 80, 91, and 119 correct last-flagged attributions, respectively.

%% file: sections/fault_extension_20260925.tex
\paragraph{Additional fault types.}
\label{app:fault-extension}
Motivated by AgentTrace's logic-error and data-corruption categories \citep{agenttrace}, we run a separate small controlled evaluation with two concrete code faults. A boundary error changes the approval condition from \texttt{risk >= 0.8} to \texttt{risk > 0.8}. A field-substitution error uses \texttt{actor\_id} instead of \texttt{owner\_id} when constructing resource-operation arguments. These are our code-level instantiations of the categories, rather than scenarios reproduced from AgentTrace. Neither adds a new request-kind trigger.

\emph{Prespecified generation.} Twelve scene attempts cover two fault families, three history types (direct transfer, multi-hop transfer, and target-local introduction), and two generation attempts per combination. Each scene contains four agents initialized from a shared action-planning function. Three receive one independent \texttt{gpt-5-mini} scheduling revision; the fourth uses a fixed valid reference implementation and independently receives the fault before the true source. Agent roles and harmless pre-introduction archive depths are randomized across scenes. Valid scheduling policies may differ on intermediate workloads; we neither require nor select for fingerprint separation. Model prompts receive synthetic task specifications and code, without source labels or the exact audit requests.

Each completed scene uses one additional model revision. Direct-transfer and target-local histories request a semantics-preserving refactor. Multi-hop histories request a benign scheduling change at the relay before retransmission, while retaining the confirmation and resource-argument behavior. Controlled fault introduction and recorded function adoptions define the origin. Validation checks that the source's parent is safe, that injection preserves all 16 online utility-case outputs, that reverting the injected edit recovers the parent, and that the recorded origin lies on the queried lineage while the independent occurrence does not. The final target version is audited. Failed constructions are retained without replacement; attribution success is not a construction criterion.

\emph{Method and evidence.} Retrieval, $K=3$, representative selection, temporal ranking, and recursive tracing remain unchanged. Each task uses its own probe and a fixed battery of 26 routine inputs, five malformed inputs, and the audit outcome. Fingerprints compare the complete returned plan, including operation arguments, so field-substitution behavior is observable. All compared methods receive the same code histories and probes; both fingerprint methods use the same signatures. Only the task-specific observations change; the source-selection algorithm remains fixed. These short controlled coding histories complement the OpenEvolve evaluation; they are not a second end-to-end evolutionary benchmark or a test of heterogeneous surrounding code.

\begin{table}[!htbp]
\centering\small
\setlength{\tabcolsep}{4pt}
\caption{Additional fault-family checks. Entries are correct-query counts; each valid scene supplies one query. Full: complete method; No rec.: multiple representatives without recursion; Exhaustive: all-version fingerprint matching without recursion.}
\label{tab:fault-extension}
\begin{tabular*}{\linewidth}{@{\extracolsep{\fill}}llrrrrr@{}}
\toprule
Fault & History & $n$ & Full & No rec. & Exhaustive & Naive CF \\
\midrule
Boundary & Direct & 2 & 2 & 2 & 2 & 2 \\
         & Multi-hop & 2 & 0 & 0 & 0 & 0 \\
         & Local & 1 & 1 & 1 & 1 & 1 \\
\addlinespace
Field substitution & Direct & 2 & 2 & 2 & 2 & 0 \\
                   & Multi-hop & 2 & 2 & 0 & 0 & 0 \\
                   & Local & 2 & 2 & 2 & 2 & 2 \\
\midrule
Overall & & 11 & 9 & 7 & 7 & 5 \\
\bottomrule
\end{tabular*}
\end{table}

\emph{Outcomes and interpretation.} Eleven of twelve attempts yield valid scenes. One boundary/local attempt fails because a generated program violates the normal scheduling specification; it is not replaced. The full method recovers 6/6 field-substitution queries and 3/5 boundary-error queries, versus 4/6 and 3/5 for exhaustive matching. Selecting the earliest unsafe external agent fails on all 11 queries because the independent occurrence is deliberately earlier. Both boundary multi-hop errors select a non-ancestor with a matching fingerprint after the relay's benign revision. A diagnostic audit of each relay's initial adoption correctly recovers the source in all four multi-hop scenes; these diagnostic queries are not included in the primary counts. Thus, the checks show transfer to additional fault operators and a concrete failure under behavioral convergence, rather than universal fault-family generalization.